\documentclass[twocolumn,prd,amssymb,amsmath,preprintnumbers,floatfix,aps,nofootinbib]{revtex4-1}
\pdfoutput=1
\usepackage{dcolumn,graphicx,bm,psfrag}
\usepackage[colorlinks=true,citecolor=blue,urlcolor=black,linktocpage=true,
linkcolor=blue]{hyperref}

\newcommand{\be}{\begin{equation}}
\newcommand{\ee}{\end{equation}}
\newcommand{\bear}{\begin{eqnarray}}
\newcommand{\eear}{\end{eqnarray}}
\newcommand{\ba}{\begin{array}}
\newcommand{\ea}{\end{array}}

\begin{document}

%
%\title{\boldmath \bf \Large Dark matter searches at neutrino experiments via monojets } 
%\title{\boldmath \bf \Large Monojets to look for light dark matter at neutrino experiments} 
%\title{\boldmath \bf \Large Probing Sub-GeV dark sectors via high energy proton beams at fixed target experiments:  MiniBooNE and LBNF} 
\title{\boldmath \bf \Large Probing sub-GeV dark sectors via  high energy  proton beams at  LBNF/DUNE  and  MiniBooNE} 

\author{\bf  Claudia Frugiuele  } 

%\vspace{.4cm}

\affiliation{Department of Particle Physics and Astrophysics, Weizmann Institute of Science, Rehovot, Israel}
%7610001
%\date{\normalsize  \small{\today}}

\begin{abstract}
We study the sensitivity to sub-GeV dark sectors of high energy ($ \geq100$ GeV) proton fixed target experiments  such as the Main Injector and the future Long-Baseline Neutrino Facility (LBNF).  We focus on off-axis detectors since they have been shown to be the ideal location to reduce the neutrino background. We consider  MiniBooNE as an off-axis detector for the NuMI facility and a hypotetical detector for LBNF located 200 m away from the target and 6.5 degrees off-axis.
We find that with the existing data, MiniBoone can explore new regions of the parameter space for leptophobic dark forces in the 100 MeV- few GeV mass range. The dedicated MiniBooNE run in beam dump mode would further improve the reach for even lighter masses. 
Therefore, MiniBooNE has the potential to be one of the most sensitive probe of leptophobic dark forces for masses between 1 MeV-10 GeV.
%We find that MiniBooNE, analyzing existing data, can already explore new regions of the parameter space for leptophobic dark forces in the 100 MeV- few GeV  mass range. 
%The dedicated  MiniBooNE run in beam dump mode could improve  even further the reach for the lighter mass region. 
%Therefore,  MiniBooNE is potentially the most sensitive probe of leptophobic dark forces in most of the 1 MeV-10 GeV parameter space.
%Therefore,  MiniBooNE can provide the strongest bounds on leptophobic dark forces in most  of the 1 MeV-10 GeV parameter space.
%Experiments using high proton beam 

% MiniBooNE is an off-axis detector repsect to the NuMI facilities and it can constrain 

%We consider the  MiniBooNE and the LBNF experiment as case example. 
%In a recent paper we showed that the ideal position for a future LBNF detector able to constraint quark-light dark matter interaction mediated by a few GeV $Z'$ would be an off-axis detector put as close as possible to the target. Interestingly we showed the  MiniBooNE/MicroBoone are already in very good position.
%It is clear that the physics case for this detector should go beyond this particular scenario. In this paper we discuss the sensitivity to such detector to mediators in the 100 MeV region.

\end{abstract}

\maketitle
\paragraph{\textbf{ Introduction}}
%The LHC is probing a broad range of energies exploring the origin of the electroweak scale, but also potentially exploring a large region of the parameter space for dark matter. It can compete with direct detection limit.
%This complementarity between direct production and direct detection is important. The LHC can be the discovery mode of dark matter.
%All these beautiful sinergy is kind of lost when we go in the light dark matter world.
%The frontier is now the sub GeV since the new generations experiments are able to probe down to 500 MeV. I can also say that the gevish is less cool!
%The two experimental evidence for physics beyond the Standard model are the 
%he existence of dark matter (DM) and the neutrino masses. 
%Similarities and difference between dark matter 
%The experimental program to shed lights on is in principle quite different blah 
The existence of dark matter (DM) is one of the most solid  pieces of evidence for physics beyond the Standard Model (SM).
We do not, however, know DM properties or if it interacts with the SM through other forces beyond gravity. Therefore, it is essential to keep an open attitude towards its investigation both from an experimental and a theoretical point of view.
In the past decade a solid experimental program for DM searches has been carried out on different levels without finding any positive signal.
In particular direct detection experiments  \cite{Cushman:2013zza} investigated the GeV-TeV scale setting impressive bounds on the quark-DM coupling.
Furthermore, the LHC is playing an important role inspecting this region of the parameter space through monojet searches 
\cite{Goodman:2010yf, Goodman:2010ku, Bai:2010hh, Fox:2011pm, Aaltonen:2012jb, Khachatryan:2014rra}. 
Both high energy colliders and direct detection experiments lose sensitivity for sub-GeV DM mass. Therefore, the question of how to look for light DM is a pressing one.
%The null results from both type of experiments suggest 
%Indirect detection also has really put in doubt the idea the there is a dark matter particle and that's it.
%Despite this we do not know anything about it, we do know 
%A broad program of search via direct detection ---> Light
%LHC
%We do not know if we are looking for the DM or if there is an extended sector to look for.
%The experimental program must be therefore as broad as possible. 
%An important question to seek answers to is whether this dark sector interact with the SM.
%In the light region
%The coupling quarks-dark  matter is more challenge to probe.

 Low energy experiments can help close this experimental loophole and a lot of effort has recently been spent studying their sensitivity \cite{Reece:2009un, Bjorken:2009mm, Essig:2010xa,Essig:2010gu, Izaguirre:2013uxa, Morrissey:2014yma, Izaguirre:2014dua, Kahn:2014sra,Gardner:2015wea,Alekhin:2015byh,Batell:2009di, deNiverville:2011it, deNiverville:2012ij, Dharmapalan:2012xp, Batell:2014yra,  Soper:2014ska,Dobrescu:2014ita,Coloma:2015pih}.
%In particular proton fixed target experiments provide an interesting experimental playground to investigate the existence  of new light degrees of freedom interacting with quarks.
%This is particularly interesting because this experiments are also the one designed to shade light on one of the other open questions of BSM physics the origin of neutrino mass.
%The idea goes as follow.
%For instance the CHARM experiment at CERN constraint 
In particular, neutrino fixed target experiments offer a way to probe light DM/quark couplings ~\cite{Batell:2009di, deNiverville:2011it, deNiverville:2012ij, Dharmapalan:2012xp, Batell:2014yra,  Soper:2014ska,Dobrescu:2014ita,Coloma:2015pih}. These experiments, aiming to investigate neutrino masses and oscillations, consist of a high-intensity proton beam impinging on a target and thus producing a large number of neutrinos via leptonic meson decays. The produced neutrino beam is then studied  via its interaction with electrons and nucleons  both in a near detector, located near the target, and in a far detector, located hundreds of km away.
% \cite{deNiverville:2011it,deNiverville:2012ij, Dharmapalan:2012xp,Batell:2014yra,Soper:2014ska,Coloma:2015pih}. 
The possibility of investigating light DM/quark interaction via this class of experiments  is particularly interesting  given the variety of present and future neutrino facilities.
%This is interesting since direct detection experiments do not have sensitivity in this region of the parameter space, while the variety of the present and future  neutrino experiments offers a good opportunity to explore it.
The idea, first proposed in \cite{Batell:2009di}, goes as follows: if we assume that the dark sector consist of a DM particle and a light mediator which also interacts with quarks, a dark matter beam is produced along the aforementioned neutrino beam. As it happens for neutrinos, these dark particles will enter the near detector and scatter with the nucleons inside.
This represents the positive and at the same time the negative side of this proposal: on one hand these experiments are accidentally designed to look  also for DM signals, but on the other hand neutrinos  represent a large background. The challenge of this program is then to suppress the neutrino background.
%The main challenge for this program is represented by the neutrino background.
In order to circumvent this problem, it was proposed in Ref.~\cite{Dharmapalan:2012xp}  to have a dedicated  run in proton beam dump mode for   MiniBooNE detector, located at Fermilab 500m away from the target striked by 8 GeV proton from the Fermilab Booster,  with the aim of suppressing the neutrino background and thus looking for light DM . This run is now complete and limits for the quark/DM mediator masses in MeV-GeV range will soon be available ( see  \cite{wc} for preliminary results).
%which will soon set strong bounds for mediator masses below the kaon threshold. 
%The clear advantage of a higher proton beam reside in the possibility to probe heavier masses for the quark/dark matter mediator, in particular the estimated maximal sensitivity for a 120 GeV beam is around 7-8 GeV,  \cite{Coloma:2015pih}.
%The DM signal inside  MiniBooNE is a neutrino- electron scattering event, while for higher proton beams 
%Experiments using higher proton beam such as the one at the Fermilab Main Injector facility or the future SHIP and Long-Baseline Neutrino Facility (LBNF) facilities ~\cite{lbnf}  could probe heavier mediator masses, for instance the Fermilab 120 GeV beam can extend the sensitivity of these experiments up to 7-8 GeV for the mediators mass.
 Higher energy proton beam ($ E \sim 100$ GeV) experiments allow to probe heavier mediator masses.
 In this regard, Ref.~\cite{Coloma:2015pih} studied the projected sensitivity to few GeV mass mediators  for Fermilab based experiments facilities  such as  the ones at the Fermilab Main Injector and the future Long-Baseline Neutrino Facility (LBNF)~\cite{lbnf}. 
This study pointed out the necessity of having an off-axis detector in order to efficiently suppress the neutrino background produced by high energy proton beams.
%Concerning other neutrino facilities, Ref.~\cite{Coloma:2015pih} studied the projected sensitivity for higher proton beam experiments such as  the ones at the Fermilab Main Injector and the future Long-Baseline Neutrino Facility (LBNF)~\cite{lbnf} . 
%A clear advantage of a higher proton beam resides in the possibility to probe heavier masses for the quark/dark matter mediator
%This study pointed out the necessity of having an off-axis detector in order to efficiently suppress the neutrino background.
 In particular, an interesting finding  is that  MiniBooNE  is in a perfect location as an off-axis detector for the NuMI beam.
 Therefore,  MiniBooNE could not only explore the light dark matter region in the dedicated beam dump mode, but could also explore the few GeV region (2 GeV-8 GeV) using existing data from the Main Injector. 
%Whereas  a clear advantage of a higher proton beam resides in the possibility to probe heavier masses for the quark/dark matter mediator,

 In this paper we investigate whether an off-axis detector could have sensitivity also to the sub-GeV region of the mediator masses. 
In particular, we want to compare the reach of  MiniBooNE as the off axis detector for the NuMI beam with the one achievable by the dedicated run in beam dump mode. This means evaluating whether a dedicated run for light DM is necessary or if the DM program can run symbiotically to the neutrino program.
Furthermore, our study  provides the necessary information to assess  and complete the physics case for  possible future off-axis detectors for LBNF/DUNE.
The paper is organized as follows: we first give a brief overview of the existing constraints on our benchmark model, a leptophobic $Z'$, with a mass in the 100 MeV-2 GeV window, which decays into DM particles and then we explain how an off-axis detector could probe this region of parameter space. The production process considered for the $Z'$ is different from the one previously studied  in the literature and the key idea is to produce DM particles from an uncollimated $Z'$ beam. After explaining this crucial point we proceed with showing the energy spectrum of DM versus the neutrino one and finally we show the sensitivity plot, where we will compare our projection to the existing constraints and to the projection for the  MiniBooNE beam dump run.
%Proton fixed target experiments provide an interesting experimental playground to probe the existence  of new light degrees of freedom interacting with quarks.
%In particular neutrino facilities   \cite{Coloma:2015pih}. offer a way to probe  the light dark matter/quarks portal. This is particularly interesting since direct detection experiments loose sensitivity in this region of the parameter space.

\paragraph{\textbf{Probing the quark/dark sector portal in the MeV-GeV region}}
\label{sect1}
As a benchmark model  we consider a sub-GeV dark matter particle, either a scalar or a fermion, charged under a new abelian gauge group  $U(1)_B$, that is:
\be
\mathcal{L}_\chi = \frac{g_z}{2}  Z^{\prime \mu} \times \left\{ 
\begin{array}{c}  z_\chi \overline \psi_\chi \gamma_\mu \psi_\chi ,   \;  
 \\ [3mm] i z_\chi \left[ (\partial_\mu \phi_\chi^\dagger) \phi_\chi  -  \phi_\chi^\dagger  \partial_\mu \phi_\chi \right] 
 \; ,  
 \end{array} \right.
\ee
where $ \psi $ is a Dirac fermion  and $ \phi$ is a complex scalar. The distinction between scalar or fermion is marginally relevant for our purpose, thus in the following we refer to the DM particle with the symbol $\chi$.
We further assume  the quarks to be  charged  under $U(1)_B$ (the simplest case is $ q_B=1/3$), while the leptons 
are neutral under it.  Models where also the leptons are charged under the new gauge group like $B-L$ or models with kinetic mixing are typically more constrained  due to the easier detectability of new degrees of freedom coupled with the electrons.
  Therefore, our benchmark model is a lepto-phobic $Z'$. 
In this paper we are interested in exploring the MeV-GeV range for an invisibly decaying $Z'$. 
  For this reason we consider  the hierarchy $ m_{\chi} <  M_{Z'}/2$ and throughout the paper we fix the DM charge $ z_{\chi}$  to be 3 in order to enforce $ BR(Z' \rightarrow \chi 
\chi) \sim 1 $. 
The present constraints on an invisibly decaying leptophobic $Z'$  in the 100 MeV- 2 GeV range are discussed in \cite{Batell:2014yra} and \cite{Dobrescu:2014ita}.
 Kaon invisible rare decays provide the strongest bound in the 100 MeV-500 MeV range.
 For heavier $Z'$  the Tevatron-monojet searches \cite{Shoemaker:2011vi} give a very mild bound constraining couplings to be $ g_z>0.1$. 
 A  slightly stronger constraint is provided  instead by $J/\Psi$ invisible decay.
  If $\chi$ is stable it provides a form of DM, see  \cite{Dobrescu:2014ita} for the discussion of cosmological bounds and relic density computation.
  It is worth mentioning that recently one of the direct detection experiments, CDMS lite \cite{Agnese:2015nto}, extended its analysis in the light DM region setting bounds for $m_{\chi} > 500 $ MeV.
  This bound is an impressive achievement, but depends crucially on the assumption that $\chi$ is the dominant dark matter particle, while the laboratory bounds have the advantage to probe a larger region of the parameter space where $\chi$ could be either a subdominant form of DM or an unstable particle decaying into lighter members of an extended dark sector.
   %Furthermore, they constrain the mass of the mediator and thus they are sensitive also to very light $\chi$ mass.
  %We will not present this bound in our final sensitivity plot both because we are typically
%Recently, CDMS lite \cite{}, a direct detection experiment, was able to push its direct detection limit down to 500 MeV for the $\chi'$ mass. Since this limit clearly depends on the assumption that $\chi$ is the dominant dark matter particle we will not present in our sensitivity plot
 %Hence, a laboratory probe in this region is still in need. 
 %and this role can be better played by proton fixed target experiments rather than by colliders.
%We can here draw a similitude between direct detection experiments and the LHC dark matter searches. This is an important complementarity, which in the low mass region it can be better played by proton fixed target experiments rather than by colliders.
Other strong bounds could arise from collider limits on  additional fermions necessary to cancel the  gauge anomalies of the $U(1)_B$ group (see \cite{Dobrescu:2014fca}).
In our sensitivity plot, which we discuss in the next section, we do not present these bounds since they are somewhat model dependent. 
% In our sensitivity plot, which we will discuss in the next section, we  do not present the indirect constraints arising from the additional fermions necessary to cancel the  gauge anomalies of the $U(1)_B$ group ( see \cite{Dobrescu:2014fca}).  The reason is they are both model dependent and indirect constraints.
  % It would be interesting to repeat the study for a scalar mediator mixed with the Higgs, but
% We further consider a scenario where the new gauge boson is the portal to a hidden sector, which for simplicity we assume to consist in a single new particle either a Dirac fermion or a complex scalar, that is:
%Our goal is  to study the potential sensitivity of off-axis detectors to sub GeV $Z'$  produced by high intensity proton beams ( $ \sim 100$) GeV.

Therefore, we see that there is an ample region of the parameter space where order one couplings are still allowed.
% MiniBooNE running in beam dump mode as proposed in \cite{Dharmapalan:2012xp}  can cover a large region of this parameter space below the kaon threshold leaving however a significant gap in the 500 MeV-few GeV range.
In the present paper we want to explore whether   MiniBooNE can explore it via the analysis of  the existing data of neutrinos coming from the NuMI beam.
%this region of the parameter space.

In Ref.\cite{Coloma:2015pih} high energy proton beams  facilities, such as LBNF and the NuMI experiments, were studied investigating the 2 GeV-8 GeV  $Z'$ mass range.
 It was shown that an off-axis location of the detectors  efficiently suppresses the background and hence sets competitive limits on $Z'$ masses in the few GeV range. This leads to the proposal of an off-axis location for a light dark matter detector and to identify 
the  MiniBooNE's location as the ideal  location to probe light dark matter  with proton beams produced by the Fermilab Main Injector.  NOvA and MINOS, instead, were shown to suffer too much from the large neutrino background due to their on-axis position.

%\textbf{mettere commento su coupling con leptons, commentare sulla massa della dm e spiegare il discorso su boosted.}
%Here we want to estimate the sensitivity of off-axis detectors to mediators masses below the GeV scale to make its physics case even stronger. 
%As in \cite{Coloma:2015pih} it was shown that  MiniBooNE with existing data would be able to set almost as strong bounds as the proposed ideal detector for LBNF, in this paper we focus on the  MiniBooNE detector used as an off-axis NuMI detector. 
% However, the statement about off-axis detectors applies more broadly; for instance SHIP can be another interesting candidate for a similar study.
\paragraph{\textbf{Detecting sub-GeV $Z'$  via off-axis detectors }}
%We want to estimate the sensitivity  to sub-GeV mediators of off-axis detectors of high proton beams.
%We want to estimate the sensitivity   to sub-GeV mediators of detectors located off-axis respect to the direction of a high energy proton beam.
%Let us summarize first the main aspects of the proposal of \cite{Coloma:2015pih} where mediators in the mass range 2 GeV-8 GeV produced via the process $ p p \rightarrow Z'$ were studied.
%Ref.\cite{Coloma:2015pih}  focused on the  2 GeV-8 GeV range  for the mediator mass and on the production via the process $ p p \rightarrow Z'$.
In the present paper we want to estimate the sensitivity of off-axis detectors to sub-GeV mediator masses. With this purpose, let us first summarize the key points of Ref. \cite{Coloma:2015pih}.
 In high energy proton beam experiments, like the one investigated, both the main neutrino and DM signal consist of neutral-current deep inelastic scattering (DIS) events, since the energy of these particles is typically a few GeV or higher.
 The challenge is finding a difference between the DM  and the neutrino events.
 DM particles are produced by the decay of  a $Z'$ mediator with mass in the 2 GeV-8 GeV range produced via $ p p \rightarrow Z'$, while neutrinos are produced via mesons  (pion and kaon) decays.
%  If the detector is placed slightly off-axis the DM energy distribution is  peaked at higher values than the neutrino background. 
%The reason is DM particles are emitted by heavier particles than mesons and this is crucial for the off-axis emission due to 
The mass difference of the parent particles is important since the energy of the emitted particle goes like:
\be
E_{\chi, \nu}=  
\frac{M^2_{Z', \pi,K}}{2 E_{Z',\pi,K} (1-\beta \cos{\theta})} .
\label{eq:Echi}
\ee
where $\theta$ is the angle between the $Z'$ and the $\chi$ direction, which for the production channel $ p p \rightarrow Z'$ corresponds to the beam direction.
Therefore, DM particles are more energetic than neutrinos.
The quantitive study of signal and  background carried in \cite{Coloma:2015pih} shows  that this difference helps in sufficiently suppressing the number of neutrino DIS events only if the detector is placed at a large angle from the target. Therefore,  the location  of  MiniBooNE  ($(\theta \sim 6.5^{ \circ}$ and 700 m away) with respect to the NuMI target is close to the ideal location,  estimated to be 200 m away from the target and $6.5^{ \circ}$ off axis. 
For sub-GeV mediators  this mass difference does not exist anymore and thus we expect  the DM energy profile to be more similar to the neutrino one. 
Therefore, one could conclude that what suppresses the neutrino background would also suppress  the signal, discarding the possibility of using off-axis detectors to constraint sub GeV mediators.
However, this is true only if DM particles are produced via the decay of a $Z'$ emitted parallel to the beam line. 
 Hence, in order to put to good use an off-axis detector to probe sub GeV $Z'$, it is necessary to consider  $Z'$ emitted off-axis with respect to the beam line.
  In this way the DM particles, reaching the  MiniBooNE detector, are emitted at a smaller angle in respect to the $Z'$ direction and therefore they are more energetic, as is clearly shown by Eq. \ref{eq:Echi}.
 % that is  the uncollimated part of the $Z'$ beam.
% The first step is to understand how we produce a dark matter beam inside this detector via light mediator.
%In this mass range we  produce a  $Z'$ beam in two ways either via meson decay  as in \cite{Batell:2014yra} either via direct production in association with a jet
%This is true for instance when the $Z'$ are produced via meson decay as in \cite{Batell:2014yra}. We expect the neutrinos to be  even less energetic than neutrinos  since they have to be emitted by $Z'$ so there is a double suppression in act.
%How do we produce a signal for light mediators inside an off-axis detector?
%WIfe need $ Z'$ produced off-axis so that dark matter particles can carry enough energy to give rise to deep inelastic scattering events.
An uncollimated $Z'$ beam is emitted either from uncollimated neutral mesons decays or via direct production of the $Z'$ together with a high $ p_T$ jet, that is via:
\begin{align}
 p + p \rightarrow  Z'+ j  \: \text {with} \:  \; p^j_T > 1 \text{GeV}
 \end{align}
 where the cut $ p^j_T>1$ GeV  guarantees the consistency of our calculation within perturbative QCD (pQCD). We will consider only this contribution, though it would  also be interesting to evaluate  the one coming from meson decays.
 %We will not take into account the contribution from meson decay.
We then simulate the process  $ p p \rightarrow Z' + j$  with $ Z \rightarrow \chi \chi $ using Madgraph \cite{Alwall:2014hca}.
For $ g_z=0.1 $ and $ N_{POT} =6 \times 10^{20} $ the number of $Z'$ produced is of order $ 10^{14}$.
 %Studying the potentially significant contribution coming from decay of  neutral mesons is outside the goal of the present paper.
%The contribution from meson decay is subleading because...
% It would be certainly interesting to take into account the meson decays contribution, but this is outside of the goals of the present paper. Therefore,  the limits presented here should be considered as conservative projections. 

\begin{figure}[t]
 \begin{center}
  \includegraphics[width=0.49\textwidth]{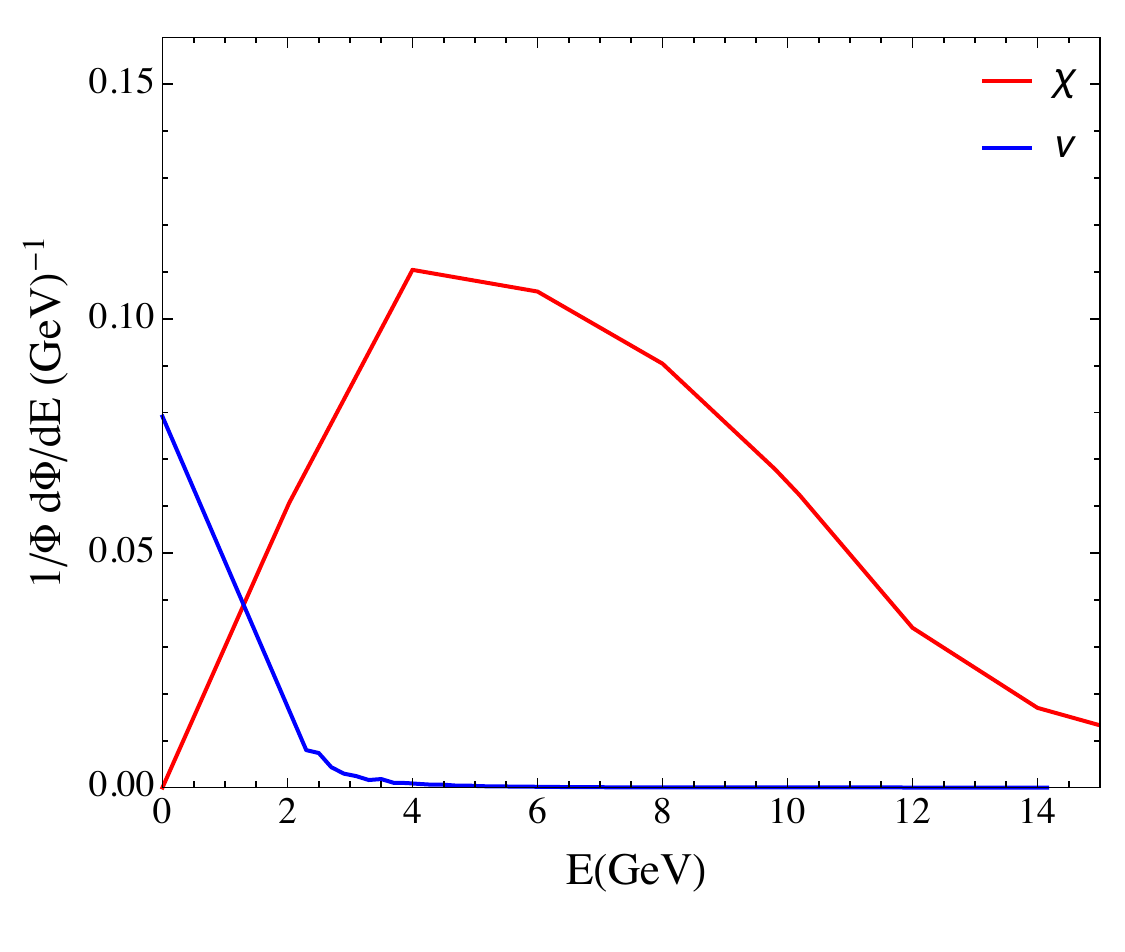} \
 \end{center}
     \caption{Energy profile of a dark  matter particle, either a complex scalar $\phi$ or a  Dirac fermion $\psi$, inside  MiniBooNE detector located 750 m away from the target and $6.5^{ \circ} $ off-axis for a benchmark point $m_{Z'}=700$ MeV and $ m_{\chi} =10$ MeV for a proton beam of 120 GeV.}
\label{fig:Edm-dist}
\end{figure}

Having established the production mechanism, the following step is studying the energy profile inside the detector of the DM particles emitted by the uncollimated $Z'$ beam.
We consider the location for the   MiniBooNE detector ( $6.5^{ \circ} $off-axis and 750 m away from the target), but our conclusion does not change for the
 proposed  dark matter LBNF ideal detector located 200 m away from the target and $6.5$ off-axis. We choose as a benchmark  a $Z'$ with mass $m_{Z'} =700$ MeV and a scalar dark matter particle with mass 10 MeV. However, the mass of the DM particle does not play a relevant role so long as it is light enough, nor its spin. 
 The energy profile for DM in this benchmark point is presented in Fig. \ref{fig:Edm-dist} together with the energy distribution for the neutrinos.
 %Fig. \ref{fig:Edm-dist} shows the energy distribution of a DM particle ( Dirac fermion solid line, complex scalar dashed) inside the ideal detector identified in  \cite{Coloma:2015pih}.
 We notice that the DM particles entering  MiniBooNE are energetic enough to give rise to deep inelastic scattering events.
 Furthermore, they are significantly more energetic than neutrinos since the $Z'$ beam is less collimated than the meson beam due to the focusing horns.
 Therefore, in order to reach the off-axis detector neutrinos need to pay the price of the large angle suppression of their energy, while DM particles are emitted at a smaller angle with respect to the off-axis direction of the $Z'$.
Having determined the energy profile of the DM beam we compute the total number of neutral current deep inelastic scattering inside MiniBooNE. The cross section for DM interacting with nucleons is much larger than the neutrino one since it is mediated by a lighter boson; for sub-GeV $Z'$ and order one coupling $ g_z \sim 0.1$ it is 1000 times bigger.
We find that the number of events is an order of magnitude bigger than the number  for heavier mediators, $ M_{Z'} >2 $GeV, due to the larger scattering cross section for the lighter mediators, which is an order of magnitude bigger than the one for few GeV $Z'$.
This difference is however tempered by the softer energy spectrum for the lighter $Z'$ case. Therefore, we do not expect a much stronger bound  on the one obtained for heavier mediators. 
%Furthermore,
 %for sufficiently light $Z'$ $( Q^2 >> M_{Z'})$,  the number of expected events become flat in mass since both the production and DIS cross section do not so the bound becomes flat as is shown below in Fig. \ref{bounds} .

%This will also provide an extra relative suppression for the background with respect to the signal, since the interaction cross section at the detector grows with the energy of the incoming particle.

\paragraph{\textbf{ Expected sensitivity }}
%We will summarize them  in a following section when comparing them with the projection of our bounds. Furthermore we will compare our projections to the one of  MiniBooNE  \cite{Batell:2014yra}.
%In the MeV-GeV mass range the strongest constraints come from rare mesons decays, in particular of pions and Kaons. In the 500 MeV- 1 GeV the only constraints is provided by mono-jets together with constraints coming from direct detection which however depend on the dark matter density. Therefore, the latter constraint our scenario only if $ \chi $ is the dominant form of dark matter.

\begin{figure}[t]
 \begin{center}
  \includegraphics[width=0.49\textwidth]{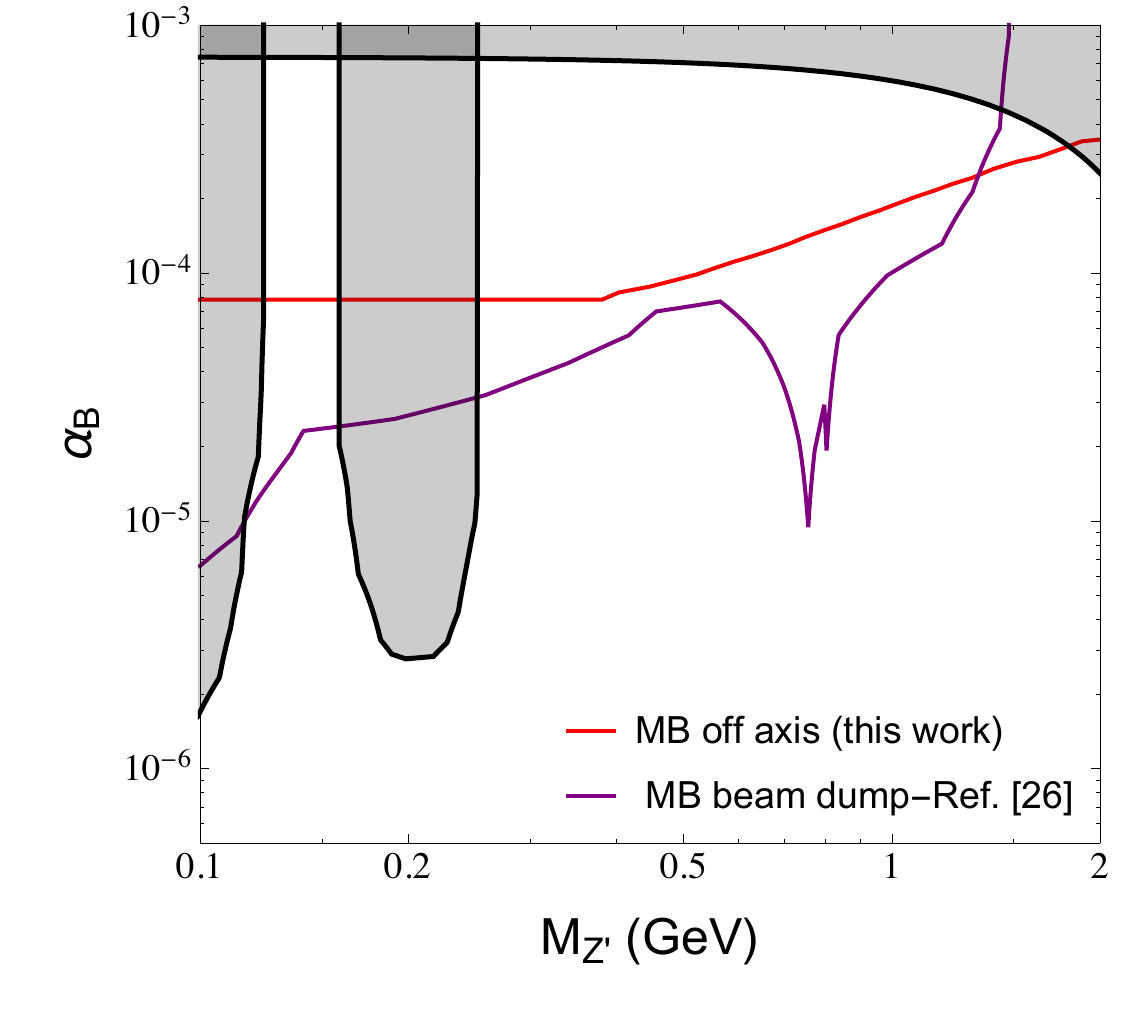} \
 \end{center}
     \caption{ Expected sensitivity (at $90 \% $C.L., 2 d.o.f.) for a fermionic dark matter particle which interacts with quarks via a  leptophobic $Z'$ in the mass/coupling plane$ (\alpha_B, M_{Z'} )$, where $ \alpha_B=  g_z^2/( 4 \pi)$. The solid red line shows the sensitivity for  a detector placed at the  MiniBooNE/MicroBooNE site. The  gray areas are ruled out ( by Kaon and $J/\Psi$ invisible decay searches), while the purple contour represents the projected sensitivity for the  MiniBooNE beam dump run  \cite{Dharmapalan:2012xp, wc}}
\label{bounds}
\end{figure}

 Fig. \ref{bounds} presents the expected sensitivity in the $ (\alpha_B, M_{Z'} )$ plane (where $ \alpha_B= g_z^2/( 4 \pi)$) for the  MiniBooNE detector computed estimating the number of  the neutrino background events coming from the Main Injector as described in \cite{Coloma:2015pih}. We do not present the projection for the LBNF ideal detector, which will be only slightly stronger due to  the smaller distance from the target. 
 As already noticed, our bound is independent from the $\chi$ spin and mass as long as $ m_{\chi} <  M_{Z'} /2$.
 We compare our projections both with the existing bounds discussed in the previous section and with the projected sensitivity of the  MiniBooNE run in beam dump mode.
Fig. \ref{bounds}  shows that  MiniBooNE analyzing the existing events coming from the Main Injector proton beam can already set the strongest bounds for masses  in a large region of the 500 MeV-2 GeV parameter space.
Therefore, combining this result with the one of  \cite{Coloma:2015pih}  MiniBooNE can set the strongest limits available on a leptophobic $Z'$ decaying into invisible particles in the region 500 MeV- 8 GeV.
However,  based  on the preliminary results presented in \cite{wc}, the beam dump mode could obtain an even better sensitivity in most of the region below 2 GeV.
%In the region below the kaon threshold our  MiniBooNE sensitivity is  almost as good as the one of the dedicated beam dump mode at least above the pion threshold.

Let us explain the difference between our proposal and the one of Ref. \cite{Dharmapalan:2012xp}. 
Our background consists of order 1000 deep inelastic neutrino events, while for the beam dump run of  \cite{Dharmapalan:2012xp}   (which uses the Booster beam line as the proton source, so an 8 GeV beam) the estimation is at most order of 100 neutron-neutrino  elastic scattering events. It might also be that  in our case a dedicated experimental analysis could further reduce our background. However, this would require different techniques than the one applied in the beam dump run since the difference in time of flight does not hold anymore for relativistic neutrinos and dark matter particles. Furthermore, the beam dump running by itself helps in lowering the background. This disadvantage is partially compensated by the fact that we have higher luminosity since our proposal is symbiotic to the neutrino program.
Finally, an important difference is played by the production mode. In our case we consider direct production of $Z'$ via pQCD processes, while in  \cite{Dharmapalan:2012xp}  the $Z'$ are produced via rare meson decays. 
%The effect of larger production production cross section is however slightly mitigated by the lower rate of interaction inside the detector compared to the rate for deep inelastic scattering. However, this increase holds also for neutrinos even if in a milder way. It would be interesting to investigate more in details the difference between high and low energy proton beams in constraining the quark/ light dark matter coupling. This is also a compelling question in order to asses the sensitivity of SHIP towards this scenario.
%Let us finally comment on the generality of our result. For instance an important question to ask is how model dependent this is.
%These experiments in general are able to constrain gauge couplings of order $ g_z \sim 0.01$; since we expect a similar sensitivity for scalar portal we immediately see that this is not enough to set competitive bounds on the Higgs portal due the Yukawa couplings suppression. On the other hand the even more suppressed coupling to the electron makes proton fixed target experiments the ideal environment to test this class of models.This provides an additional reason to improve the constraining power of these facilities towards invisibly decaying mediators.
\paragraph{\textbf{Conclusions}}
Several fixed  target experiments such as LBNF and SHIP will be running in the coming years and many other experiments are currently in full action.  It is therefore important to try to find ways to use the potential of these experimental devices. In the present paper we investigated the sensitivity to sub-GeV dark/visible sector mediators  of   MiniBooNE as off-axis detectors for the NuMI beam finding that:
\begin{itemize}
\item  Combining our result together with the finding of Ref.\cite{Coloma:2015pih}  we conclude that  MiniBooNE is the most sensitive probe of the 500 MeV-8 GeV region for a leptophobic $Z'$ which decays invisibly. This will be possible already using existing data  and does not require an additional run. This is symbiotic to the neutrino program.
\item  In a large part of the sub GeV region our limits are almost comparable, but somewhat weaker than the projected limits for the dedicated run in beam dump mode. For lighter mediators  mass this additional run provides limits significantly stronger than our sensitivity. Therefore,  our analysis is complementary to \cite{Dharmapalan:2012xp} and performing it would allow  MiniBooNE to set the strongest available bounds on dark leptophobic forces in most of the 1 MeV-10 GeV region.
\item Our analysis offers interesting  and general perspectives for high energy proton fixed target experiments. It would be interesting to do a similar study for SHIP since a new background analysis  would be necessary, but this we postpone to a later publication.
\end{itemize}
%\section*{  Comments ( not part of the draft)}

%We continue with this idea of adopting collider tool to the fixed target/low intensity frontier.
%I fix here it would be cool to make a sort of analogy to monojet search
%Capire il discorso su produzione inclusiva o esclusiva ( per i neutrini io voglio produrre pions, Zprime plus jets invece nn me ne frega nulla).
%Inoltre devo dimostrare che la produzione Z off shell and jet nel caso di 
%SHIP:Capire anche se il fatto di avere tau neutrini può' risultare in un problema/ Ora ho mesoni più pesanti che producono il mio bkg. Cmq posso dire che lo studio lo postpongo dopo 
%\newline
%Sicuro una delle cose su cui dobbiamo insistere di più che noi stiamo proponendo una nuova analisi usando vecchi dati di  MiniBooNE
%Mi chiedo se uno possa dire qualcosa se guardi troppe NC rispetto a quante CC osservi, tipo NC/CC e capire
%Una cosa su cui voglio riflettere e' il fatto che la massa in questo nn entra, ma entra in direct detection. Commentare su questo punto.
%%%%%%%%%%%%%%%%%%%%%%%%%%
\paragraph*{Acknowledgements} 
I would like to thank Pilar Coloma for collaboration in the early stage of the project and Tim Tait for insightful discussions. I am grateful to Kfir Blum, Marco Farina, Roni Harnik, Yossi Nir and Lorenzo Ubaldi for interesting comments and careful reading of the manuscript.
 %Part of this work was carried out  at the Aspen Center for Physics, which is supported by National Science Foundation grant PHY-1066293.

 \bibliography{DMneut}

\end{document}